\documentclass[apj]{emulateapj}
\usepackage{mathptmx}

\def\gtorder{\mathrel{\raise.3ex\hbox{$>$}\mkern-14mu
             \lower0.6ex\hbox{$\sim$}}}
\def\ltorder{\mathrel{\raise.3ex\hbox{$<$}\mkern-14mu
             \lower0.6ex\hbox{$\sim$}}}

\slugcomment{Submitted to ApJL}

\shorttitle{{\em Kepler} and the Oort cloud}

\shortauthors{Ofek \& Nakar}

\begin{document}

\title{Detectability of Oort cloud objects using {\em Kepler}}
\author{Eran~O.~Ofek\altaffilmark{1}$^{,}$\altaffilmark{2} and Ehud~Nakar\altaffilmark{3}}

\altaffiltext{1}{Division of Physics, Mathematics and Astronomy, California Institute of Technology, Pasadena, CA 91125, USA}
\altaffiltext{2}{Einstein Fellow}
\altaffiltext{3}{School of Physics and Astronomy, Tel-Aviv University, Tel-Aviv 69978, Israel}

\begin{abstract}

The size distribution and total mass of objects in the Oort Cloud
have important implications to the theory of planets formation,
including the properties of, and the processes taking place in the
early solar system.
%
We discuss the potential of space missions like {\em Kepler}
and {\em CoRoT},
designed to discover transiting exo-planets,
to detect Oort Cloud, Kuiper Belt and main belt objects
by occultations of background stars. Relying
on published dynamical estimates of the content of the Oort Cloud,
we find that {\em Kepler}'s main program is expected to detect between $0$ and
$\sim100$ occultation events by deca-kilometer-sized Oort Cloud
objects.
The occultations rate depends on the mass of the Oort cloud,
the distance to its ``inner edge'', and the size distribution of its objects.
In contrast, {\em Kepler} is unlikely to find occultations by Kuiper
Belt or main belt asteroids, mainly due to the fact that it is
observing a high ecliptic latitude field.
Occultations by Solar System objects will appear
as a photometric deviation in a single measurement,
implying that the information regarding the
time scale and light-curve shape of each event is lost.
We present statistical methods that have the potential
to verify the authenticity of occultation events by Solar
System objects, to estimate the distance to the occulting population,
and to constrain their size distribution.
Our results are useful for planning of future
space-based exo-planet
searches in a way that will
maximize the probability of detecting solar system objects, without
hampering the main science goals.

\end{abstract}


\section{Introduction}
\label{sec:Intro}

Oort (1950) postulated
the existence of a cloud of comets orbiting the Sun with typical semi-major
axes, $a$, of $\sim10^{4}$\,AU.
This cloud is required to explain
the existence of long period comets
and the fact that a considerable fraction of these comets have
orbital energies
concentrated in a narrow range,
corresponding to $1/a\approx 10^{-4}$\, AU$^{-1}$.
However, to date no direct observation of objects in the Oort Cloud
exist\footnote{It is possible that (90377) Sedna, 2000\,CR105 and 2006\,SQ372 belong to the inner Oort Cloud.}.

Dynamical simulations suggest that the Oort Cloud was formed by the
ejection of icy planetesimals from the Jupiter-Neptune region by
planetary perturbations (e.g., Duncan et al 1987; Dones et al.
2004). On orbital time scales, galactic tides and passage of nearby
stars raised the perihelia of these comets above the region of the
giant planets influence.
%
Comets with semi-major axes above $\sim2\times10^{4}$\,AU are loosely bound to the Sun,
and tides by the galactic potential (e.g., Heisler \& Tremaine 1986)
as well as impulse by passing nearby stars,
may send these comets to the inner solar system where they reveal themselves as
long period comets.
In contrast, Oort Cloud objects with $a\ltorder 2\times10^{4}$\,AU
(``inner Oort Cloud''; e.g., Hills 1981, Bailey 1983)
have more stable orbits.
We note that the inner ``boundary'' of the Oort Cloud, $r_{{\rm min}}$, is set
by perturbations due to the giant planets
and the primordial stellar neighborhood of the Sun.
This $r_{{\rm min}}$ is estimated to be
in the range of 1,000\,AU to 3,000\,AU
(Duncan et al. 1987; Dones et al. 2004; Fern\'{a}ndez 1997).
Below $r_{{\rm min}}$
the number of comets steeply falls due to planetary influence.
Dynamical simulations suggest that the spatial density of Oort Cloud
objects in the 3,000--50,000\,AU range
falls as $r^{\alpha}$,
with $\alpha=-3.5\pm0.5$
(e.g., Duncan et al. 1987).

The current flux of long period comets calibrated by dynamical simulations
suggest that the outer Oort Cloud (i.e., $a>2\times10^{4}$\,AU)
contains $\sim10^{12}$ comets with nuclei absolute planetary magnitude\footnote{Defined
as the magnitude of an object observed at opposition and
at 1\,AU from the Sun and Earth.}
brighter than 16\,mag (e.g., Heisler 1990; Weissman 1996). This magnitude roughly corresponds to objects with radius of 2\,km (assuming $4\%$ geometric albedo).
The dynamically more stable inner Oort Cloud
contains about 2--100 more comets than its outer counterpart
(e.g., Hills 1981; Duncan et al. 1987; Fern\'{a}ndez \& Brunini 2000; Dones et al. 2004;
Brasser et al. 2008).

Detecting Oort Cloud objects,
and estimating their size distribution, will improve our understanding
of the mass of the solar-system planetary accretion disk,
and the dynamical processes in the young solar system.
Moreover, it may 
bear some clues to the stellar density at the region in which
the Sun was born (e.g., Fern\'{a}ndez 1997; Fern\'{a}ndez \& Brunini
2000; Brasser et al. 2006).
However, these objects are beyond the direct reach of even
the currently largest planned telescopes,
and some indirect methods (e.g., astrometric microlensing; Gaudi \& Bloom 2005).

Bailey (1976)
suggested that small Kuiper Belt objects (KBOs)
as well as Oort Cloud objects
could be detected by occultations of background stars.
Indeed several surveys are looking for KBO occultations
(e.g.,
Roques et al. 2006; Lehner et al. 2009;
Chang et al. 2006, 2007;
Bianco et al. 2009; Schlichting et al. 2009).
However, to date there is only one reported
occultation of a Kuiper Belt object (Schlichting et al. 2009).

In this letter we
discuss the potential of space missions designed to search for
transiting exo-planets (e.g., {\em Kepler}, {\em CoRoT}) 
to detect occultations of background stars by small
Oort Cloud and Kuiper Belt objects.

\section{Stellar occultations by small solar system objects}
\label{Analytic}

Space telescope designed to look for exo-planet transits
usually have exposure times, $t_{{\rm exp}}$,
longer by orders of magnitude than the typical
duration of an occultation of a background star by
a small solar system object, $t_{{\rm dur}}$.
The relative flux decrement
of such an occultation $\Delta$,
integrated over the exposure time,
is diluted by a factor $\propto t_{{\rm dur}}/t_{{\rm exp}}$.
However, these satellites have
superb photometric accuracy.
Their noise level per exposure
is $\sigma = S t_{{\rm exp}}^{-1/2} \sim 10^{-4}-10^{-5}$,
where $S$ is a normalization
parameter\footnote{For {\em Kepler} ({\em CoRoT})
$S=0.0022$ ($0.045$) for 1\,s integration of a 12th magnitude star.
This sensitivity parameter depends on the star magnitude $m$ as
$10^{+0.2(m-12)}$.}.
Therefore, we
may be able to detect these minute occultations.
Here, we adopt a detection criterion
for an occultation of $\Delta > n\sigma$,
where $n$ is the signal-to-noise ratio.
Throughout the paper we adopt $n=8$.

For an occultation by objects at a given distance from Earth, $r$,
the maximal value of $\Delta$ (obtained for an optimal impact
parameter\footnote{In the case of diffraction
occultation it is not at $b=0$.} $b$) increases
monotonically with the object size.
We define $R_{{\rm min}}$ as the radius of the smallest
object that can be detected, and correspondingly
$\theta_{{\rm min}}=R_{{\rm min}}/r$.
Furthermore, we can define  the sum of
impact parameters for which objects 
with a radius $R$ are detectable, $b_{\Sigma}(R,n\sigma)
\equiv 2 \int_{0}^{\infty} \Theta[\Delta(R,b)-n\sigma]db$, where
$\Theta$ is Heaviside step function. According to these definitions
$b_{\Sigma}(R<R_{{\rm min}},n\sigma)=0$.

The size distribution of KBOs and Oort Cloud objects per unit solid angle
is parametrized by a power-law,
$dN(R)/dR = N_{>1\,{\rm km}}(1\,{\rm km})^{-1} (q-1) (R/1\,{\rm
km})^{-q}$,
or by a broken power-law.
Here, $N_{>1\,{\rm km}}$ is the total number of objects
with $R>1$\,km per unit area,
and $q$ is the size distribution power-law index. For Oort
Cloud objects $AN_{>1\,{\rm km}}\sim 10^{13}$,
where $A$ is the surface area of the celestial sphere.
For KBOs larger than $R_{{\rm break}}\approx 45$\,km $q=4.5$
(Bernstein et al. 2004; Fuentes \& Holman 2008; Fraser et al. 2008),
and
$q\approx 3-4$ for smaller KBOs
(e.g, Farinella \& Davis 1996; Pan \& Sari 2005; Schlichting et al. 2009).
For Oort Cloud objects $q$ is unconstrained by observations
(see however Goldreich et al. 2004).

The number of events in a survey that observes $N_*$ stars for a
duration $\tau$ is then
$N_{{\rm ev}} \cong N_* \tau \int_{R_{{\rm min}}}^{\infty} \mu b_{\Sigma}(R,n\sigma) \frac{dN}{dR} dR$,
where $\mu$ is the angular
velocity of the occulting objects on the sky as seen from Earth.
Here all objects are at the same distance $r$, all stars
have the same magnitude and angular radii
and $\mu$ is constant. When this is not the case, it is
straightforward to integrate over the distribution of these
properties.

The functional forms of $t_{\rm dur}$, $\Delta$, $R_{\rm min}$ and
thus $b_{\Sigma}(R)$ depend on the ratios between the three
angular scales in the problem. These are the object angular radius,
$\theta_{{\rm obj}}=R/r$; the stellar angular
radius\footnote{$\theta_{*}\propto T_{{\rm e}}^{-2}10^{-0.2 M_{{\rm
bol}}}$, where $T_{{\rm e}}$ and $M_{{\rm bol}}$ are the effective
temperature and bolometric magnitude of the star, respectively.}
$\theta_*$; and
the angular Fresnel scale\footnote{For $r=3000$\,AU and
$\lambda=5000$\,\AA,~F=10.6\,km and $\theta_{{\rm
F}}=4.9\times10^{-6}$ arcsec.} $\theta_{{\rm F}}=F/r$, with
$F=\sqrt{\lambda r /2}$, and $\lambda$ the wavelength of
observation. The duration of the eclipse is $t_{{\rm dur}}\approx
2\theta_{{\rm max}}/ \mu$, where $\theta_{{\rm max}}={\rm max}
(\theta_{{\rm obj}},\theta_{{\rm F}},\theta_{*})$.
Table~\ref{Tab:Rmin} gives the functional form of
$R_{\rm min}$ in the different ``asymptotic'' cases (denoted by
A--E), where the parameters are normalized to those of a typical
Oort Cloud object and the {\em Kepler} sensitivity.

The cases where $\theta_{{\rm obj}} \gg \theta_*, \theta_{{\rm F}}$ (case A)
and $\theta_* \gg \theta_{{\rm obj}}, \theta_{{\rm F}}$ (cases B--C) represent
geometric occultations. In these cases we approximate
$b_{\Sigma}(R) \cong \max{ \{ \theta_{{\rm obj}}, \theta_{*} \}} (11.7-q)/13$
for any $R>R_{{\rm min}}$.
Here, the $q$-dependent factor (i.e., $[11.7-q]/13$)
is a result of
the assumed circular shape of the object.
We obtained this approximate correction factor by numerically
integrating geometrical occultations by circular shaped
objects\footnote{This integral do~not have an analytical solution.}.
In the diffractive regime ($\theta_{{\rm F}} \gtorder \theta_{*},\theta_{{\rm obj}}$;
Cases D--E) we numerically calculated diffractive light curves
using Eq.~B1 in Roques et al. (1987). We find that $b_{\Sigma}$ can
be well approximated by
\begin{equation}
b_{\Sigma}(R,n\sigma) \cong
\frac{\theta_{{\rm F}}^2}{\max\{\theta_{{\rm obj}},\theta_*\}}
                           \left[1-\frac{n\sigma}{4} \frac{t_{{\rm exp}}}{t_{{\rm dur}}}
                                   \left(\frac{\theta_{{\rm F}}}{\theta_{{\rm obj}}}\right)^2
                           \right].
\label{Eq:DeltaDiffractive}
\end{equation}
The scaling of Eq. \ref{Eq:DeltaDiffractive} can be understood as
follows. The occultation ``shadow'' pattern of an object can be
described as a set of bright and ``dark'' concentric rings with the
central ring (around the ``Poisson peak'') being dark with an
angular radius of $\theta_{{\rm F}}$. The area in all the rings is
constant ($\sim \theta_{{\rm F}}^2$) so the angular width of the rings is
inversely proportional to its angular distance from the center (e.g.,
Fig.~1 in Nihei et al 2007). The 'depth' of the decrement in the
flux in the dark rings is $\approx ( \theta_{{\rm obj}}/\theta_{{\rm F}})^{2}$
so the value of $\Delta$ for the central ring occultation is
$\Delta_{{\rm c}} \sim (t_{{\rm dur}}/t_{{\rm exp}}) ( \theta_{{\rm obj}} / \theta_{{\rm F}} )^{2}$.
This decrement
remains similar in all the 'dark' rings as long as the ring width is
smaller than $\max\{\theta_*,\theta_{{\rm obj}}\}$. Therefore
$b_{\Sigma}(R,n\sigma \lesssim \Delta_{{\rm c}}) \sim \theta_{{\rm F}}^2/\max{
\{\theta_{*},\theta_{{\rm obj}} \} }$.

For each of the five cases we use the approximations described above
and give formula (Table~\ref{Tab:Nev})
for the expected number of events as a function of
the parameters $n$, $S$, $t_{{\rm exp}}$, $\theta_{*}$, $r$, $\lambda$ and
$v_{{\rm rel}}~(=\mu r)$.
Here $v_{{\rm rel}}$ is the relative velocity
between the occulting object and Earth, projected on the plane
perpendicular to the Earth-object direction.
Since the object size
distribution is steep, the integration over $R$ is dominated by
the smallest objects, and is carried to infinity (rather than the
radius at which the condition defining the ``case'' is violated).

\begin{deluxetable*}{llll@{}l@{}l@{}l@{}l@{}l}
\tablecolumns{9}
\tablewidth{0pt}
\tablecaption{Occultation regimes and the Minimum radius of detectable objects}
\tablehead{
\colhead{Case} &
\colhead{Condition} &
\colhead{$R_{{\rm min}}\cong$} &
\colhead{[km]} &
\colhead{} &
\colhead{} &
\colhead{} &
\colhead{}
}
\startdata
A & $\theta_{{\rm obj}}\gg \theta_{*}, \theta_{{\rm F}}$
  & $ 7.9$
  &            $\frac{n}{8} \frac{S}{0.0022} $
  &            $\Big( \frac{t_{{\rm exp}}}{900\,{\rm s}} \Big)^{1/2}$
  &            $\frac{v_{{\rm rel}}}{30\,{\rm km\,s}^{-1}}$
  &
  &
  & \\

B & $\theta_{*}\gg\theta_{{\rm obj}}\gg\theta_{{\rm F}}$
  & $ 16.6$
  &             $\Big( \frac{n}{8} \frac{S}{0.0022} \Big)^{1/2}$
  &             $\Big( \frac{t_{{\rm exp}}}{900\,{\rm s}} \Big)^{1/4}$
  &             $\Big( \frac{v_{{\rm rel}}}{30\,{\rm km\,s}^{-1}} \Big)^{1/2}$
  &             $\Big( \frac{r}{3000\,{\rm AU}} \Big)^{1/2}$
  &             $\Big( \frac{\theta_{*}}{1.6\times10^{-5}\,{\rm ''}} \Big)^{1/2}$
  & \\
C & $\theta_{*}\gg\theta_{{\rm F}}\gg\theta_{{\rm obj}}$
  & $ 14.0$
  &             $\Big( \frac{n}{8} \frac{S}{0.0022} \Big)^{1/2}$
  &             $\Big( \frac{t_{{\rm exp}}}{900\,{\rm s}} \Big)^{1/4}$
  &             $\Big( \frac{v_{{\rm rel}}}{30\,{\rm km\,s}^{-1}} \Big)^{1/2}$
  &             $\Big( \frac{r}{3000\,{\rm AU}} \Big)^{1/2}$
  &             $\Big( \frac{\theta_{*}}{1.6\times10^{-5}\,{\rm ''}} \Big)^{1/2}$
  & \\
D & $\theta_{{\rm F}}\gg\theta_{{\rm obj}}\gg\theta_{*}$
  & $ 4.6$
  &            $ \Big( \frac{n}{8} \frac{S}{0.0022} \Big)^{1/2}$
  &            $ \Big( \frac{t_{{\rm exp}}}{900\,{\rm s}} \Big)^{1/4}$
  &            $ \Big( \frac{v_{{\rm rel}}}{30\,{\rm km\,s}^{-1}} \Big)^{1/2}$
  &            $ \Big( \frac{r}{3000\,{\rm AU}} \Big)^{1/4}$
  &
  &            $ \Big( \frac{\lambda}{5000\,{\rm \AA}} \Big)^{1/4}$ \\
E & $\theta_{{\rm F}} \gg \theta_{*} \gg \theta_{{\rm obj}}$
  & $ 4.6$
  &            $ \Big( \frac{n}{8} \frac{S}{0.0022} \Big)^{1/2}$
  &            $ \Big( \frac{t_{{\rm exp}}}{900\,{\rm s}} \Big)^{1/4}$
  &            $ \Big( \frac{v_{{\rm rel}}}{30\,{\rm km\,s}^{-1}} \Big)^{1/2}$
  &            $ \Big( \frac{r}{3000\,{\rm AU}} \Big)^{1/4}$
  &
  &            $ \Big( \frac{\lambda}{5000\,{\rm \AA}} \Big)^{1/4}$
\enddata
\tablecomments{The normalization of the
star angular radius, $\theta_{*}$, corresponds to $R=35$\,km at $r=3000$\,AU.
The difference between case B and case C
is that in case C we replace $\theta_{{\rm obj}}$ by $2^{1/4}\theta_{{\rm obj}}$.
This small correction is required in order to take into
account the increased cross section of diffractive occultations
(estimated numerically).
All the formulas assume that $t_{{\rm exp}}>t_{{\rm dur}}$.}
\label{Tab:Rmin}
\end{deluxetable*}
\begin{deluxetable*}{ll@{}l@{}l@{}l@{}l@{}l@{}l}
\tablecolumns{8}
\tablewidth{0pt}
\tablecaption{Formulas for the expected number of occultation events in the different asymptotic regimes}
\tablehead{
\colhead{Case} &
\colhead{$N_{{\rm ev}}\approx$}
\colhead{} &
\colhead{} &
\colhead{} &
\colhead{} &
\colhead{} &
\colhead{} 
}
\startdata
A &
    $ 50 \times 10^{0.90(3.5-q)}
                       \frac{11.7-q}{13}
                       \frac{q-1}{q-2}$
  &
                       $\frac{AN_{>1\,{\rm km}}}{10^{13}}
                       \frac{N_{*}}{10^{5}}
                       \frac{\tau}{3\,{\rm yr}}
                       \Big( \frac{n}{8} \frac{S}{0.0022} \Big)^{2-q}$
  &
                       $\Big( \frac{t_{{\rm exp}}}{900\,{\rm s}} \Big)^{\frac{2-q}{2}}$
  &
                       $\Big( \frac{v_{{\rm rel}}}{30\,{\rm km\,s}^{-1}} \Big)^{3-q}$
  &
                       $\Big( \frac{r}{3000\,{\rm AU}} \Big)^{-2}$
  &
  & \\
B &
    $ 35 \times 10^{1.22(3.5-q)}
                       \frac{11.7-q}{13}$
  &
                       $\frac{AN_{>1\,{\rm km}}}{10^{13}}
                       \frac{N_{*}}{10^{5}}
                       \frac{\tau}{3\,{\rm yr}}
                       \Big( \frac{n}{8} \frac{S}{0.0022} \Big)^{\frac{1-q}{2}}$
  &
                       $\Big( \frac{t_{{\rm exp}}}{900\,{\rm s}} \Big)^{\frac{1-q}{4}}$
  &
                       $\Big( \frac{v_{{\rm rel}}}{30\,{\rm km\,s}^{-1}} \Big)^{\frac{3-q}{2}}$
  &
                       $\Big( \frac{r}{3000\,{\rm AU}} \Big)^{-\frac{1+q}{2}}$
  &
                       $\Big( \frac{\theta_{*}}{1.6\times10^{-5}\,{\rm ''}} \Big)^{\frac{3-q}{2}}$ 
  & \\
D &
    $  350 \times 10^{0.66(3.5-q)}
                       \frac{3.5(q-1)}{q(q+2)}$
  &
                       $\frac{AN_{>1\,{\rm km}}}{10^{13}}
                       \frac{N_{*}}{10^{5}}
                       \frac{\tau}{3\,{\rm yr}}
                       \Big( \frac{n}{8} \frac{S}{0.0022} \Big)^{-\frac{q}{2}}$
  &
                       $\Big( \frac{t_{{\rm exp}}}{900\,{\rm s}} \Big)^{-q/4}$
  &
                       $\Big( \frac{v_{{\rm rel}}}{30\,{\rm km\,s}^{-1}} \Big)^{\frac{2-q}{2}}$
  &
                       $\Big( \frac{r}{3000\,{\rm AU}} \Big)^{-\frac{q+4}{4}}$
  &
  &
                       $\Big( \frac{\lambda}{5000\,{\rm \AA}} \Big)^{\frac{4-q}{4}}$ \\
E &
    $  36 \times 10^{0.66(3.5-q)}
                       \frac{4.5}{q+1}$
  &
                       $\frac{AN_{>1\,{\rm km}}}{10^{13}}
                       \frac{N_{*}}{10^{5}}
                       \frac{\tau}{3\,{\rm yr}}
                       \Big( \frac{n}{8} \frac{S}{0.0022} \Big)^{\frac{1-q}{2}}$
  &
                       $\Big( \frac{t_{{\rm exp}}}{900\,{\rm s}} \Big)^{\frac{1-q}{4}}$
  &
                       $\Big( \frac{v_{{\rm rel}}}{30\,{\rm km\,s}^{-1}} \Big)^{\frac{3-q}{2}}$
  &
                       $\Big( \frac{r}{3000\,{\rm AU}} \Big)^{-\frac{7+q}{4}}$
  &
                       $\Big( \frac{\theta_{*}}{1.6\times10^{-5}\,{\rm ''}} \Big)^{-1}$
  &
                       $\Big( \frac{\lambda}{5000\,{\rm \AA}} \Big)^{\frac{5-q}{4}}$
\enddata
\tablecomments{Case C is like case B but multiplied by
$(1/2)^{(1-q)/4}$.
$A$ is the area of the celestial sphere.
Note that these formulas are
not continuous and
they provide a better approximation in the ``asymptotic'' cases.
All the formulas assume that $t_{{\rm exp}}>t_{{\rm dur}}$.}
\label{Tab:Nev}
\end{deluxetable*}

\section{Occultation rates for {\em Kepler} and {\em CoRoT}}
\label{rate}

To estimate the observed rate of events
we need to take into account
the radial distance distribution (discussed in \S\ref{sec:Intro}),
the relative velocity ($v_{{\rm rel}}$) probability distribution,
the ecliptic latitude distribution,
and the stellar angular size and magnitude co-distributions.

\subsection{Stellar angular size and magnitude distribution}

\begin{figure}
\centerline{\includegraphics[width=8.5cm]{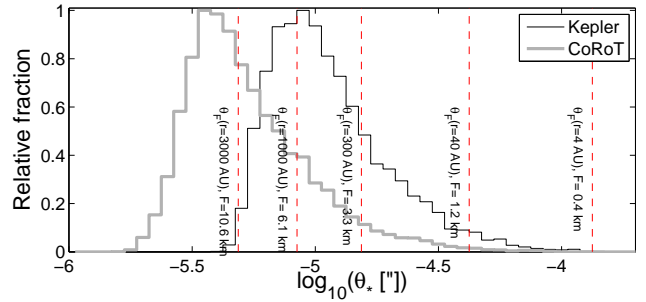}}
\caption{Approximate stellar angular radius distributions for
{\em Kepler} ({\it black} line) and {\em CoRoT} ({\it gray} line).
The vertical {\it red dashed} lines represent the Fresnel radii at selected distances (see labels) where $\lambda=5000$\,\AA.
\label{fig:Sat_StarsAngRadDist}}
\end{figure}

In Figure~\ref{fig:Sat_StarsAngRadDist} we present the estimated
angular radii distribution of target stars observed
by {\em Kepler} and {\em CoRoT}.
To estimate this
we queried the guide star catalog
(GSC; version 2.2; Lasker et al. 2008)
for all stars within 2\,deg from the center of the {\em Kepler} ({\em CoRoT})
field, that have
$R$-band magnitudes between 9 and 14 (12 and 16), and $B-R$ colors between
0.59 and 3.4\,mag.
This color range roughly corresponds to the spectral range of
{\em Kepler}
targets.
We converted the color index of each star to an effective temperature
and a bolometric correction
by fitting the color indices with those obtained from synthetic
photometry of black-body spectra with different temperatures
(e.g., Ofek 2008).
The synthetic photometry (Poznanski et al. 2002)
was preformed using the transmission curves
of the filters used in the GSC (Moro \& Munari 2000).

\subsection{Relative velocity distribution}

Given the unit vector in the direction of a target $\hat{r}$,
the speed of the observer relative to the occulting object $v$,
and the heliocentric velocity unit vector of the
observer\footnote{Object velocity is neglected.}
$\hat{v}$,
the approximate relative velocity,
of an observer in an Earth trailing orbit,
projected on the plane perpendicular
to $\hat{r}$ is
\begin{equation}
v_{{\rm rel}} = v\sqrt{1 - (\hat{v} \cdot \hat{r})^{2}} \cong
          v_{\oplus}\sqrt{1 - \cos^{2}{(\lambda-\lambda_{\oplus})}\cos^{2}{\beta}},
\label{Eq:Vrel}
\end{equation}
where $\lambda$ and $\beta$ are the ecliptic longitude and latitude
of the observed star,
$\lambda_{\oplus}$ is the ecliptic longitude
(as measured from the Sun)
of the observer,
and $v_{\oplus}\approx 29.8$\,km\,s$^{-1}$ is the heliocentric
speed of the observer.
%
Finally, assuming circular orbit
(i.e., the probability distribution of $\lambda_{\oplus}$ is uniform)
the probability
distribution
of $v_{{\rm rel}}$
is $P_{v_{{\rm rel}}} = \frac{d\lambda_{\oplus}}{dv_{{\rm rel}}}$.

\subsection{Ecliptic latitude distribution}

The Oort Cloud is predicted to have a roughly spherical
shape.
However, the Kuiper Belt and the asteroids main belt
are concentrated toward the ecliptic plane.
Using the inclination distribution of KBOs estimated by
Elliot et al. (2005) the expectation probability to
find a KBO at an ecliptic latitude of $\gtorder 66^{\circ}$ is about
$f_{\epsilon}\sim 6\times 10^{-6}$
times smaller than that at the ecliptic.
%
%
This relative probability was estimated by convolving the inclination
probability distribution of KBOs
with the ecliptic latitude distribution
for a given inclination (i.e., Elliot et al. 2005 eqs.~40 and 34, respectively).
We note, however, that this number is highly uncertain.

\subsection{Rate estimates}


To estimate the rate of occultation events
we integrate the appropriate formulas in Table~\ref{Tab:Nev}
over $v_{{\rm rel}}$, $r$, $\theta_{*}$ and $S$ co-distributions.
For each combination of parameters we integrate
over the most appropriate occultation channel (i.e., cases A--E).
In Table~\ref{Tab:Rates} we present the estimated rates predicted
for {\em Kepler} and {\em CoRoT} for the Oort Cloud (upper block),
Kuiper Belt (middle block), and main belt (lower block) object
occultations.
The predicted rate for Oort Cloud objects in Table~\ref{Tab:Rates} in case of
$q=3.5$ and $r_{{\rm min}}=3000$\,AU is lower by about an order of
magnitude compared to the pre-factors in
Table~\ref{Tab:Nev}.
This is mostly because Table~\ref{Tab:Nev} is normalized
for all objects being at $r=3000$ AU while in
Table~\ref{Tab:Rates} the same number of objects is spread between
$r_{{\rm min}}$ and $r_{{\rm max}}$, with a spatial density distribution
proportional to $r^{-3.5}$.

%
%
\begin{deluxetable}{lllcc}
\tablecolumns{5}
\tablewidth{0pt}
\tablecaption{Approximate occultation rates}
\tablehead{
\multicolumn{3}{c}{Parameters} &
\multicolumn{2}{c}{Rates ($\tau=3$\,yr)} \\
\colhead{} &
\colhead{} &
\colhead{} &
\colhead{{\em Kepler}} &
\colhead{{\em CoRoT}} \\
\colhead{$N_{>1\,{\rm km}}$} &
\colhead{$r_{{\rm min}}$} &
\colhead{$q$} &
\colhead{$f_{\epsilon}=6\times10^{-6}$} &
\colhead{$f_{\epsilon}=1$} \\
\colhead{[deg$^{-2}$]} &
\colhead{[AU]} &
\colhead{} &
\colhead{$N_{*}=10^{5}$, $\beta\approx66^{\circ}$} &
\colhead{$N_{*}=10^{4}$, $\beta\approx 0^{\circ}$}
}
\startdata
$2.4\times10^{8}$                      & $3000$ & $3.0$ & 30           &  0.1               \\
$2.4\times10^{8}$                      & $3000$ & $3.5$ &  8           &  $4\times10^{-3}$   \\
$2.4\times10^{8}$                      & $3000$ & $4.5$ & 0.7          &  $1\times10^{-5}$   \\
$2.4\times10^{8}$                      & $1000$ & $3.5$ & 60           &  0.03              \\
$2.4\times10^{8}$                      & $5000$ & $3.5$ & 3            &  $2\times10^{-3}$   \\
\hline
$1.1\times10^{4} f_{\epsilon}$ & $40$   & $3.0$\tablenotemark{a} & $2\times10^{-4}$     & $3\times10^{-3}$    \\
$7.3\times10^{4} f_{\epsilon}$ & $40$   & $3.5$ & $5\times10^{-4}$     & $4\times10^{-3}$    \\
$4.9\times10^{5} f_{\epsilon}$ & $40$   & $4.0$ & $5\times10^{-3}$     & $0.02$              \\
\hline
$47 f_{\epsilon}$              & $4$    & $2.3$\tablenotemark{b} &$1\times10^{-4}$      & 0.2
\enddata
\tablenotetext{a}{For KBOs we assume a broken power-law with $q=4.5$ above R=45\,km. The actual
rate calculation is preformed by normalizing the Equations in Table~\ref{Tab:Nev}
to 45\,km, and using different $q$ for $R_{{\rm min}}$ above and below 45\,km.}
\tablenotetext{b}{For main belt asteroids we assume a broken power-law with $q=4$ above R=2.5\,km,
and $q=2.3$ below this radius.}
\tablecomments{The first block refers to Oort Cloud objects
assuming $r_{{\rm max}}=50000$\,AU, $\alpha=-3.5$.
The second block refers to KBOs assuming $r_{{\rm max}}=40$\,AU.
The third block refers to main belt asteroids assuming $r_{{\rm max}}=4$\,AU.
%
The cumulative surface density of KBOs near the break radius,
$R_{{\rm break}}\approx 45$\,km,
is 5.4\,deg$^{-2}$ at the ecliptic (Fuentes et al. 2009).
Therefore, at the ecliptic we adopt for KBOs,
$N_{>1\,{\rm km}}$ of
$1.1\times10^{4}$,
$7.3\times10^{4}$,
and $4.9\times10^{5}$\,deg$^{-2}$,
for $q=$3, 3.5, and 4, respectively.
We note that these values are consistent with the findings of
Schlichting et al. (2009).
$N_{{>1\,\rm km}}$ and $q$ for the main belt are taken from Ivezi\'{c} et al.
(2001).
We set the ecliptic latitude surface density correction
$f_{\epsilon}$ to $6\times10^{-6}$ and 1 for
$\beta\approx66^{\circ}$ and $\beta\approx0$, respectively, for both
KBOs and main belt asteroids.
In all the cases we assume eight sigma detection threshold (i.e., $n=8$).}
\label{Tab:Rates}
\end{deluxetable}

These rate estimates suggest that if our current
'best guess' regarding the properties of the Oort Cloud
are correct, then it may be detectable by {\em Kepler}.
Regardless, {\em Kepler} will be able to put
the best constraint (so far) on the content of the
Oort Cloud.
We note that the rate estimates in Table~\ref{Tab:Rates}
are an approximation to the actual rate.
This is mainly because our formulas represent asymptotic
approximations, while in some cases two out of the three angular scales,
$\theta_{{\rm obj}}$, $\theta_{{\rm F}}$, $\theta_{*}$,
maybe be similar.

\section{Validation and degeneracy removal}

Contrary to transits and grazing transits by extra-solar planets, solar system
occultations will appear in a single photometric measurement and
they happen only once. Therefore, verification of each event by
itself will require a thorough understanding of the noise
properties of the photometer.
{\em Kepler} will have about $1.5\times10^{10}$ 15-min independent
photometric measurements. Assuming purely Gaussian noise,
the 8-$\sigma$ detection threshold we adopted
corresponds to a probability of $\sim10^{-15}$.
However, even small deviations from a Gaussian noise may
impair our confidence in the astrophysical nature of the events.

Here we present two tests that can be used to check if the occultations
signature are due to real events or due to some sort of uncharacterized noise.
In addition, these tests can be used to statistically measure
$q$ and to distinguish between Oort Cloud objects and KBOs.


The occultation rate depends on the relative velocity $v_{{\rm rel}}$,
with a somewhat different dependency in each regime.
Therefore,
if a large number of occultations are detected,
a simple sanity test is to look for the
dependency of the events rate on $v_{{\rm rel}}$.
We note that since {\em Kepler} is observing at high ecliptic latitude
the $v_{{\rm rel}}$ modulation will be small (Eq.~\ref{Eq:Vrel})
compared to the maximal modulation which will
be seen by an instrument observing the ecliptic.
Moreover, in all the cases (A--E), the function $N_{{\rm ev}}$
depends on $q$.
Therefore, the amplitude of the modulation 
is related to $q$ and the occultation type (i.e., A--E).
Knowing the occultation regime (see below),
we can measure $q$ (i.e., removing the degeneracy
between $q$ and $N_{{\rm ev}}$).

Furthermore, as shown in Table~\ref{Tab:Nev}
the occultations rate depends on $S$, which depends on the
stellar magnitude and on $\theta_{*}$, which in turn is a function
of the star magnitude and effective temperature.
We expect bright and angularly small stars to
have higher chance of producing detectable occultations.
Moreover, this dependency is different for the various occultation cases.
This can be used to further test the nature of the occultations,
by comparing the distribution of magnitudes and effective temperatures of the
stars showing occultations to those of the entire
observed stellar population.
Most importantly,
for a {\em Kepler}-like observatory,
KBO occultations will be mostly in the diffractive regime,
while Oort Cloud object occultations will be in the
geometric regime (Fig.~\ref{fig:Sat_StarsAngRadDist}).
Therefore, the magnitude and $\theta_{*}$ of occulted
stars by KBOs and Oort Cloud objects
will follow different co-distributions.
Therefore, we can statistically
distinguish between Oort Cloud and KBO occultation events.
However, quantifying these effects and estimating the
number of occultations needed to carry out the suggested analysis
requires a more realistic estimate of the events rate
(i.e., using full numerical integration of light curves)
and a full exploration of the parameters space.

To summarize, we show that {\em Kepler} and similar space missions
can detect deca-kilometer objects in the Oort Cloud.
We present statistical methods to verify that the occultations
are real rather than due to uncharacterized noise.
In addition we suggest that it will be possible to
statistically measure
their size distribution ($q$),
and the dominant occultation regime, and therefore differentiate between
Oort cloud objects and KBOs.

Considering {\em Kepler} capabilities and reasonable Oort Cloud
parameters, we find that {\em Kepler} may detect 0 to $\sim10^{2}$
stellar occultations by Oort Cloud objects.
We find that {\em Kepler} is unlikely to detect Kuiper Belt
objects (mainly because of its high ecliptic latitude pointing).
Moreover, {\em CoRoT} is unlikely to detect Kuiper Belt or Oort Cloud
objects. However, the exact properties of the Oort Cloud and the
Kuiper Belt are not well known. Therefore, such searches are
warranted.

The analytical treatment we present in this paper is useful in order
to maximize the efficiency of existing (and future, e.g., {\em
PLATO}) experiments to detect trans Neptunian objects. However,
this treatment is accurate only in the asymptotic
cases, and numerical calculations are
needed in order to give more precise predictions.

\acknowledgments
We thank Re'em Sari, Hilke Schlichting, Orly Gnat and Michael Busch
for many valuable discussions.
Support for program number HST-AR-11766.01-A was provided by NASA.
EOO is supported by an Einstein fellowship.
EN is partially supported by an IRG grant.

\end{document}